\documentclass[preprint,review,12pt]{elsarticle}

\usepackage{amssymb}

\journal{Physica E}

\begin{document}

\begin{frontmatter}



\title{Anisotropic resistivity of the monolayer graphene in the trigonal warping and connected Fermi curve regimes}


\author{J. Azizi}
\author{A. Phirouznia\corref{cor1}}
\cortext[cor1]{Corresponding author.} \ead{Phirouznia@azaruniv.ac.ir}
\author{K. Hasanirokh}
\address{Department of Physics, Azarbaijan Shahid Madani
              University, 53714-161, Tabriz, Iran}

\begin{abstract}
In the present study, the anisotropic resistivity of the monolayer graphene has been obtained in semiclassical regime beyond the Dirac point approximation. In particular, detailed
investigations were made on the dependence of conductivity on the Fermi energy. At low energies, in the vicinity of the Dirac points, band energy of the monolayer graphene is isotropic at the Fermi level. Meanwhile, at the intermediate Fermi energies anisotropic effects such as trigonal warping is expected to be the origin of the anisotropic resistivity. However, besides the band anisotropy there also exists an other source of anisotropic resistivity which was introduced by scattering matrix. At high energies it was shown that the band anisotropy is less effective than the anisotropy generated by the scattering matrix. It was also shown that there exist two distinct regimes of anisotropic resistivity corresponding the trigonal warping and connected Fermi curve at intermediate and high energies respectively.
\end{abstract}


\end{frontmatter}


\section{Introduction}
\label{intro}
Two-dimensional crystals of carbon atoms (a single sheet of graphite) \cite{1} was first fabricated in 2004 by Novoselov et. al. \cite{2}. Discovery and fabrication of graphene provided a matchless opportunity for novel experimental observation of electronic transport properties which has also provided a rich field of theoretical studies over the last ten years. The experimental realization of a graphene has prompted much excitement and emotion in both the experimental and theoretical physics. From a fundamental point of view, discovery of graphene was important not only in providing the first realization of Dirac Hamiltonian and relativistic massless particles \cite{14,15,16,17,18} but also in providing a way for designing graphene-based electronic devices. The discovery of these extraordinary properties in graphene-based systems in recent years opens unprecedented expectancy for the investigation of low dimensional systems.
\\
The energy bands of graphene touch together in the edge of the hexagonal Brillouin zone, known as Dirac points. Energy spectrum of carriers is linear at the Dirac points. This fact has many significant consequences especially on the electric transport in graphene. Therefore the electrical transport in graphene becomes very active research field in recent years because of its potential application in nano-material and instrumentation of nano-scale materials. It should be noted that it was shown that the graphene based nano-structures such as nano-ribbons could have finite energy gap at the Dirac points \cite{33,35,36,38,39,40,41,42,44,45,46,47,48,Chen}.
\\
As mentioned before in the pure graphene the energy band in the edge of the hexagonal Brillouin zone meet each other. This fact provides a theoretical perception to realize unusual transport properties in this material. Graphene is a gapless semiconductor with a minimal conductivity which can be considered as the nearly universal value of the order of $4e^2/h$ \cite{23,24,25,26,27,28}. Meanwhile this conductivity depends on externally imposed conditions such as the temperature and doping. The band structure of the graphene has been obtained in the 1947 by Wallace \cite{10} however, the universal value of the minimal conductivity in the pure graphene is not completely understood until the recent years.
\\
In the present work it was shown that the Fermi energy, $\epsilon_F$, determines different transport regimes. Unlike the linear energy dispersion at low energies (typically when $0<\epsilon_F\leq 1 eV$) in the vicinity of the Dirac points, small corrections, such as second order of Dirac equation, \cite{11} would lead to revision in effective Hamiltonian of graphene at higher energies. These corrections which appears in the energy dispersion by introducing an additional quadratic term results in deformation of the Fermi line. The deformation of the Fermi
circle around a K-point in which the circular Fermi curve at the Dirac points changes to a trigonal known as trigonal warping.
In fact breaking the symmetry of the effective Hamiltonian at the Dirac points results in trigonal warping \cite{11,13,29,30}. This effect has been reported in graphene-related structures such as bilayer and multi-layer graphene and even in carbon nano-tubes \cite{trig1,trig2,trig3}.
It was also shown that, by increasing the Fermi energy beyond the hopping energy, $t$, another regime will appear in which the shape of the Fermi curves and the behavior of the anisotropic resistivity (AR) are changed simultaneously.
\section{Model}
Anisotropic transport generally has been discussed in term of the asymmetry of the scattering between two states on the Fermi surface. In the present work we have employed an analytical approach which was introduced by V\'{y}born\'{y} and et al in \cite{12}. They have described an analytical approach in which the anisotropic transport can be obtained within the semiclassical Boltzmann method \cite{12}. If we consider the Boltzmann equation for non-equilibrium distribution function, $f_\lambda(\vec{k},\varepsilon )$, as
\begin{eqnarray}
-e\vec{\varepsilon} .\vec{v}_\lambda(\vec{k})(-{{\partial }_{\epsilon }}{{f}_{0}})=\sum_{\lambda'}\int{\frac{{{d}^{2}}{{k}^{'}}}{{{(2\pi )}^{2}}}}\omega_{\lambda\lambda'} (\vec{k},{{\vec{k}}^{'}})[f_\lambda(\vec{k},\varepsilon )-f_{\lambda'}({{\vec{k'}}},{{\varepsilon }})],
\end{eqnarray}
where $\omega_{\lambda\lambda'}(\vec{k},\vec{k}')$ denotes the scattering rate between the following states: $|\vec{k}\lambda>$ and $|\vec{k'}\lambda'>$. In this approach the following solution has been proposed for non-equilibrium distribution function
\begin{eqnarray}
f_\lambda(\phi ,\theta )-{{f}_{0}}=-e\varepsilon v_\lambda(-{{\partial }_{\epsilon }}{{f}_{0 \lambda}})(a_\lambda(\phi )\cos (\theta)+b_\lambda(\phi )\sin (\theta ))
\end{eqnarray}
Where $\phi$ and $\theta$ are angles along the $\vec{k}$ (wave vector) and $\vec{\varepsilon}$ (electrical field ) respectively. $\lambda$ and $\lambda'$ are the band index, $f_0$ is the equilibrium distribution function and the velocity, $\vec{v}_\lambda$,
$(\vec{v}_\lambda=\frac{1}{\hbar }{{\nabla }_{k}}{{\epsilon }_{\vec{k}}})$ is given by the band dispersion energy. The electric field and wave vector have been denoted by ${\vec{\varepsilon}}=\varepsilon (\cos \theta ,\sin \theta ), {\vec{k}}=k(\cos \phi ,\sin \phi )$ respectively.
\\
By writing the Taylor series of the distribution function
\begin{eqnarray}
f(\vec{k},\vec{\varepsilon} )={{f}_{0}}+{{\varepsilon }_{x}}{{\partial }_{{{\varepsilon }_{x}}}}f+{{\varepsilon }_{y}}{{\partial }_{{{\varepsilon }_{y}}}}f+\sum{{{\varepsilon }_{i}}{{\varepsilon }_{j}}{{\partial }_{{{\varepsilon }_{i}}}}{{\partial }_{{{\varepsilon }_{j}}}}f}+...
\end{eqnarray}
For a two band system ($\lambda=\pm$) by using the above equations it can be shown that in order to have the non-equilibrium distribution function following relations have to be satisfied \cite{12}.
\begin{eqnarray}
\label{main1}
\cos (\phi )&=&{{\overline{\omega} }_{\pm}}(\phi ){{a}_{\pm}}(\phi )\\&&-{{\int{d{{\phi }^{'}}[{{\omega }_{\pm\pm}}(\phi ,\phi }}^{'}}){{a}_{\pm}}({{\phi }^{'}})+{{\omega }_{\pm\mp}}(\phi ,{{\phi }^{'}}){{a}_{\mp}}(\phi )] \nonumber
\end{eqnarray}
\begin{eqnarray}
\label{main2}
\sin (\phi )&=&{{\overline{\omega} }_{\pm}}(\phi ){{b}_{\pm}}(\phi )\\&&-{{\int{d{{\phi }^{'}}[{{\omega }_{\pm\pm}}(\phi ,\phi }}^{'}}){{b}_{\pm}}({{\phi }^{'}})+{{\omega }_{\pm\mp}}(\phi ,{{\phi }^{'}}){{b}_{\mp}}(\phi )] \nonumber
\end{eqnarray}
In which we have assumed
\begin{eqnarray}
\overline{\omega }_{\lambda}(\phi )&=&\sum_{\lambda'}\int{d{{\phi' }}\omega_{\lambda\lambda'}  (\phi ,{{\phi'}})}\\
\omega_{\lambda\lambda'} (\phi ,{{\phi' }})&=&{{(2 \pi )}^{-2}}\int{{{k'}}d{{k'}}\omega_{\lambda\lambda'}  (k,{{k'}})}
\end{eqnarray}
Where $a_{\pm}(\phi)$ and $b_{\pm}(\phi)$ take the form of the Fourier series that can be describe by
\begin{eqnarray}
a_{\pm}\,(\phi \,)\,&=&\,{{a}_{0 }}\,+\,{{a}^{\pm}_{c1}}\,cos\,\phi \,+\,{{a}^{\pm}_{c2}}\,cos2\phi \,+\,\ldots \nonumber\\
&&+\,{{a}^{\pm}_{s1 }}\,sin\,\phi \,+\,{{a}^{\pm}_{s2 }}\,sin2\,\phi \,+\,\ldots  \\
b_{\pm}(\phi \,)\,&=&\,{{b}_{0 }}\,+\,{{b}^{\pm}_{c1 }}\,cos\,\phi \,+\,{{b}^{\pm}_{c2}}\,cos2\phi \,+\,\ldots  \nonumber\\
 &&+\,{{b}^{\pm}_{s1}}\,sin\,\phi \,+\,{{b}^{\pm}_{s2 }}\,sin2\,\phi \,+\,\ldots
\end{eqnarray}
\\
Provided that the coefficients ${a}_{\pm}$ and ${b}_{\pm}$ are known by solving the equations (\ref{main1})-(\ref{main2}) the non-equilibrium distribution functions are given for each band as follows
\begin{eqnarray}
{{f}_{+}}(\phi ,\theta )-{{f}_{0+}}&=&-e\varepsilon v_+(-{{\partial }_{\epsilon }}{{f}_{0 +}})\left[ {{a}_{+}} cos \theta +{{b}_{+}}(\phi )sin\theta  \right] \nonumber\\
{{f}_{-}}(\phi ,\theta )-{{f}_{0-}}&=&-e\varepsilon v_-(-{{\partial }_{\epsilon }}{{f}_{0- }})\left[ {{a}_{-}}(\phi )cos \theta +{{b}_{-}}(\phi ) sin\theta  \right]\nonumber\\
\end{eqnarray}
\section{Anisotropic conductivity beyond the Dirac point}
Tight binding Hamiltonian of pure graphene in the nearest neighbor approximation is given by
\begin{eqnarray}
H_0=-t\sum_{<i,j> }\,({{a}^{\dagger }}_{i}{{b}_{j}}+h.c)
\end{eqnarray}
In which the operators $a_{i}^{\dagger }$ and ${{b}_{j}}$ refer to the creation and annihilation of an electron in sublattices A and B respectively and $t=2.7$ eV denotes the hopping parameter.
\\
Matrix representation of the Hamiltonian in the bases $\psi
=({{\psi }_{A }},
   {{\psi }_{B}})$ is as follows
\\
\begin{eqnarray}
\,{{H_0}}=\left(
\begin{array}{ll}
 0 ~~& H_{AB}(\textbf{k})  \\
 H_{AB}^*(\textbf{k}) &  ~~0 \\
 \end{array}
\right),
\end{eqnarray}
where $H_{AB}(\textbf{k})
=t(e^{-i\textbf{k}.{{{\vec{\delta}}}_{1}}}+e^{-i\textbf{k}.{{{\vec{\delta}}}_{2}}}+e^{-i\textbf{k}.{{{\vec{\delta}}}_{3}}})$ and we have defined nearest neighbors position vectors by
${{\vec{\delta}}_{1}}=\frac{a}{2}(1,\sqrt{3}),\,{{\vec{\delta}}_{2}}=\frac{a}{2}(-1,\sqrt{3}),\,{{\vec{\delta}}_{3}}=a(-1,0)$ in which
the carbon-carbon distance is denoted by $a=1.42${\AA}.
\\
The eigen-states may then be written as
\begin{eqnarray}
\psi_k=\frac{1}{\sqrt{2}}\left(
\begin{array}{c}
   \lambda{{e}^{i\bar{\varphi }_k}}  \\
   1  \\
\end{array} \right),
\end{eqnarray}
In which
\begin{eqnarray}
\bar{\varphi_k}(\phi)={{\tan }^{-1}}(\frac{{Im}H_{AB}(\vec{k})}{{Re}H_{AB}(\vec{k})}),\\
\end{eqnarray}
and
\begin{eqnarray}
{\phi}&=&{{\tan }^{-1}}(\frac{k_y}{k_x}),\\
\end{eqnarray}
Then the band energies are given by
\begin{eqnarray}
\epsilon_{k\lambda}&=&\lambda t (1+4 \cos(3 a/2k_x )\cos( \sqrt{3}a/2k_y)\nonumber\\
                   &&+4\cos^2( \sqrt{3}a/2)k_y)^{1/2}
\end{eqnarray}
Where $\lambda=\pm 1$ is the band index.
Unlike the Dirac point Hamiltonian here the energy spectrum is anisotropic in k-space.
\\
In the presence of the impurities the hamiltonian of the system reads
\begin{eqnarray}
H=H_0+V_{im}.
\end{eqnarray}
In which $V_{im}(\vec{r})=v\sum_j\delta(\vec{r}-\vec{r}_j)$ stands for short range impurity potential in which summation is over the position of the impurities and $v$ is the strength of the impurity potential. The scattering rates are defined through the relations:
\begin{eqnarray}
{{\omega }_{++}}(\phi,\phi')=\frac{\pi}{\hbar}n_i({{v}^{2}}+{{v}^{2}}\cos (\bar{\varphi_k }-\bar{\varphi_{k'} })))\nonumber \\
{{\omega }_{--}}(\phi,\phi')=\frac{\pi}{\hbar}n_i({{v}^{2}}+{{v}^{2}}\cos (\bar{\varphi_k }-\bar{\varphi_{k'} }))) \\
{{\omega }_{+-}}(\phi,\phi')=\frac{\pi}{\hbar}n_i({{v}^{2}}-{{v}^{2}}\cos (\bar{\varphi_k }-\bar{\varphi_{k'} })))\nonumber \\
{{\omega }_{-+}}(\phi,\phi')=\frac{\pi}{\hbar}n_i({{v}^{2}}-{{v}^{2}}\cos (\bar{\varphi_k }-\bar{\varphi_{k'} }))).\nonumber
\end{eqnarray}
Where $n_i$ is the density of the impurities. Scattering rates can be expressed in a compact form as follows
\begin{eqnarray}
{{\omega }_{\lambda\lambda'}}(\phi,\phi')=\frac{2\pi}{\hbar}n_i{v}^{2}F_{\lambda\lambda'}(kk').
\end{eqnarray}
In which $F_{\lambda\lambda'}(kk')=({1}+\lambda\lambda'\cos (\bar{\varphi_k }-\bar{\varphi_{k'} }))/2$ is the form factor of the given states $|k\lambda>$ and $|k'\lambda'>$.
\\
A two-dimensional Fourier decomposition of the scattering rates has been performed in order to figure out the equations for $a_\lambda$ and $b_\lambda$. Then we can write
\begin{eqnarray}
w_{\lambda\lambda'}(\phi ,{{\phi }^{'}})&=&{{A}^{\lambda\lambda'}_{0}}+\sum_{mn}{A^{\lambda\lambda'}_{mn}\cos(m\phi +n{{\phi }^{'}})}\nonumber\\
&&+\sum_{mn}{B^{\lambda\lambda'}_{mn}\sin (m\phi +n{{\phi }^{'}})}
\end{eqnarray}
If the above Fourier expansion has been continued up to $m,n\leq N$. In this case the equations (\ref{main1})-(\ref{main2}) results in $8\times N$ linear equations which can be described by the following linear matrix equations
\begin{eqnarray}
\textbf{M}\left(\begin{array}{c}
{\textbf{{\textit{a}}}_{c}}^{+}  \\
{\textbf{{\textit{a}}}_{s}}^{+}  \\
{\textbf{{\textit{a}}}_{c}}^{-}  \\
{\textbf{{\textit{a}}}_{s}}^{+}  \\
\end{array}\right)=
\left(\begin{array}{c}
   \textbf{1}_{N\times 1}  \\
   \textbf{0}_{N\times 1}  \\
   \textbf{1}_{N\times 1}  \\
   \textbf{0}_{N\times 1}  \end{array}\right)
   \\
\textbf{M}'\left(\begin{array}{c}
{\textbf{{\textit{b}}}_{s}}^{+}  \\
{\textbf{{\textit{b}}}_{c}}^{+}  \\
{\textbf{{\textit{b}}}_{s}}^{-}  \\
{\textbf{{\textit{b}}}_{c}}^{-}  \\
\end{array}\right)=
\left(\begin{array}{c}
   \textbf{1}_{N\times 1}  \\
   \textbf{0}_{N\times 1}  \\
   \textbf{1}_{N\times 1}  \\
   \textbf{0}_{N\times 1}  \end{array}\right).
\end{eqnarray}
In which
\begin{eqnarray}
&&\textbf{M}=
\\
\nonumber\\
&&\left(
\begin{array}{cccc}
\gamma_1 I-\pi {\textbf{A}}^{++}&  -\pi {{\textbf{A}}}^{+-}         & -\pi {\textbf{B}}^{++}         & -\pi{{\textbf{B}}}^{+-} \\
-\pi {{\textbf{A}}}^{-+}         &  \gamma_2 I-\pi {{\textbf{A}}}^{--}&  -\pi {{\textbf{B}}}^{-+}        &-\pi {{\textbf{B}}}^{--}     \\
-\pi {{\textbf{B}}}^{++}          &  -\pi {{\textbf{B}}}^{+-}         & \gamma_1 I+\pi {{\textbf{A}}}^{++}&  \pi {{\textbf{A}}}^{+-} \\
-\pi {{\textbf{B}}}^{-+}          & -\pi {{\textbf{B}}}^{--}         &\pi {{\textbf{A}}}^{-+}         &{{\gamma }_{2}} I+\pi{{\textbf{A}}}^{--} \\
\end{array} \right),\nonumber\\
\nonumber\\
&&\textbf{M}'=
\\
\nonumber\\
&&\left(
\begin{array}{cccc}
\gamma_1 I+\pi {\textbf{A}}^{++}&  \pi {{\textbf{A}}}^{+-}         & -\pi {\textbf{B}}^{++}         & -\pi{{\textbf{B}}}^{+-} \\
\pi {{\textbf{A}}}^{-+}         &  \gamma_2 I+\pi {{\textbf{A}}}^{--}&  -\pi {{\textbf{B}}}^{-+}        &-\pi {{\textbf{B}}}^{--}     \\
-\pi {{\textbf{B}}}^{++}          &  -\pi {{\textbf{B}}}^{+-}         & \gamma_1 I-\pi {{\textbf{A}}}^{++}&  -\pi {{\textbf{A}}}^{+-} \\
-\pi {{\textbf{B}}}^{-+}          & -\pi {{\textbf{B}}}^{--}         &-\pi {{\textbf{A}}}^{-+}         &{{\gamma }_{2}} I-\pi{{\textbf{A}}}^{--} \\
\end{array} \right),\nonumber\\
\end{eqnarray}
where $\textbf{A}^{\lambda\lambda'}=[A_{mn}^{\lambda\lambda'}]$ and $\textbf{B}^{\lambda\lambda'}=[B_{mn}^{\lambda\lambda'}]$ are $N \times N$ matrices of the Fourier coefficients, $\gamma_1 =\pi(A_0^{++}+A_0^{-+})$, ${{\gamma }_{2}}=\pi(A_0^{--}+A_0^{+-})$ and $I$ is a $N\times N$ unit
matrix. Meanwhile the unknown coefficients of the distribution function are given by
\begin{eqnarray}
{\textbf{{\textit{a}}}^{\pm}_{\eta}}=\left(\begin{array}{c}
{{a}^{\pm}_{\eta 1}}  \\
{{a}^{\pm}_{\eta 2}}  \\
{{a}^{\pm}_{\eta 3}}  \\
\vdots                  \\
{{a}^{\pm}_{\eta N}}  \\
\end{array}\right)_{N\times 1}
   \\
{\textbf{{\textit{b}}}^{\pm}_{\eta}}=\left(\begin{array}{c}
{{b}^{\pm}_{\eta 1}}  \\
{{b}^{\pm}_{\eta 2}}  \\
{{b}^{\pm}_{\eta 3}}  \\
\vdots                  \\
{{b}^{\pm}_{\eta N}}  \\
\end{array}\right)_{N\times 1},
\end{eqnarray}
in which $\eta=s$ or $c$.
\\
Once ${\textbf{{\textit{a}}}^{\pm}_{\eta}}$ and ${\textbf{{\textit{b}}}^{\pm}_{\eta}}$ are determined, by solving the given linear equations, we can obtain the non-equilibrium distribution function for each band.
This can be achieved by inserting the obtained coefficients $a_{\eta}(\phi)$ and $b_{\eta}(\phi)$ in
$f_\lambda(\phi ,\theta )={f}_{0}-e\varepsilon v_\lambda(-{{\partial }_{\epsilon }}{{f}_{0 \lambda}})(a_\lambda(\phi )\cos (\theta )+b_\lambda(\phi )\sin (\theta ))$. Then current and conductivity of the sample are given by
\begin{eqnarray}
\vec{j}(\varepsilon,\theta)&=&\sum_\lambda\int\frac{d^2k}{(2\pi)^2}e\vec{v}_\lambda(\vec{k})f_\lambda(\phi ,\theta )\nonumber\\
\sigma_{xx}&=&{j}(\varepsilon,\theta=0)/\varepsilon\\
\sigma_{yy}&=&{j}(\varepsilon,\theta=\pi/2)/\varepsilon.\nonumber
\end{eqnarray}
And finally the anisotropic resistivity of the system
can be defined as
\begin{eqnarray}
AR=-\frac{\sigma_{xx}-\sigma_{yy}}{\sigma_{xx}+\sigma_{yy}}
\end{eqnarray}
\section{Anisotropic resistivity at Dirac points}
Most of the interesting physical properties of graphene are the manifestations of the linear energy dispersion relation at the Dirac points. The Hamiltonian of the gapped graphene can be expressed by the following expression at the Dirac point
\begin{eqnarray}
H=-i\hbar v_F(\sigma_x\partial_x+\sigma_y\partial_y)+\alpha {{\sigma }_{z}}
\end{eqnarray}
$\alpha$ is the amount of the energy gap, $\sigma_i$ is the Pauli matrix representing pseudospin degree of freedom and $v_f$ is Fermi velocity. The corresponding
eigenvectors can be easily obtained as follows:
\begin{eqnarray}
\mid \psi_{\pm}(k)>=\left(
\begin{array}{cc}
{{m}_{\pm }}\,{{e}^{-i\phi }}\\
 \,\,\,\,\,\,1\,\,\, \\
\end{array}\right)\,{e}^{ik.r},
\end{eqnarray}
where
${{m}_{\pm }}=(\alpha \,\pm \,\sqrt{{{\alpha }^{2}}\,+{{(\hbar \,k\,{{v }_{f}})}^{2}}})/\hbar k{{v }_{f}}$. Using the forgoing approach it can be easily shown that $AR=0$ i.e. at low Fermi energies transport in graphene is direction-free and absolutely isotropic.
\section{RESULT AND DISCUSSION}
\begin{figure}
  \includegraphics{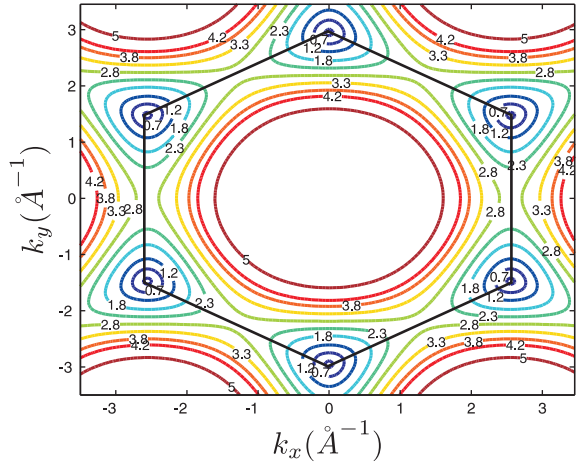}
\caption{Fermi curves of the monolayer graphene at different Fermi energies. Each band has been labeled by the value of corresponding Fermi energy in term of eV. When $0\leq \epsilon_F<t$ Fermi curves appear as distinct circles or deformed triangles. Meanwhile at higher Fermi energies ($t<\epsilon_F$) Fermi curve appears as a continues and connected curve. }
\label{fig1}       
\end{figure}

\begin{figure*}
  \includegraphics[width=0.95\textwidth]{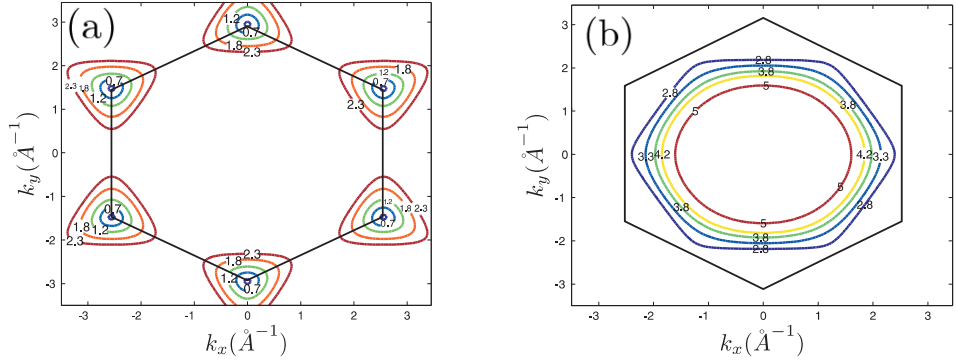}
\caption{Fermi curves of the monolayer graphene at (a) low ($\epsilon_F<t$) and (b) high ($\epsilon_F>t$) Fermi energies. Low Fermi energies correspond to the Dirac and trigonal warping regimes (a). High Fermi energies correspond to the connected Fermi curve regime (b). }
\label{fig2}       
\end{figure*}

\begin{figure}
  \includegraphics{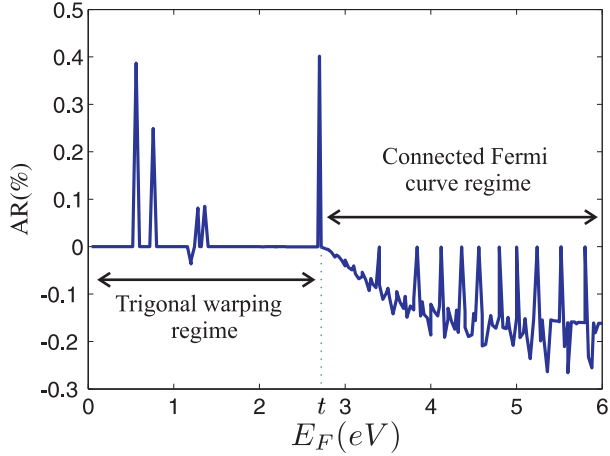}
\caption{Anisotropic resistivity as a function of the Fermi energy. There are two distinct regimes of the anisotropic resistivity which are identified by the value of the Fermi energy ($\epsilon_F$) with respect to the hopping amplitude ($t$).  }
\label{fig3}       
\end{figure}

We take typical parameters such as $v=0.4$ eV for the strength of the impurities and $t=2.7$ eV for hopping amplitude. Accuracy of the numerical solution of the linear equations directly depends on the number of the terms in $a_\eta(\phi)$ and $b_\eta(\phi)$ expansions. We have shown that good convergence between the different results can be obtained when $N>6$. In the current case we take $N=7$ which results in $56$ linear equations.
\\
The real-space hexagonal symmetry in graphene reflects itself in k-space as well. This type of the discrete symmetry clearly removes the spatial isotropy along the $x$ and $y$ directions. Therefore anisotropic response function has been expected when the external electric field is directed along these directions. Since the band anisotropy changes at different Fermi energies it is expected that the anisotropy of the response function should be a function of Fermi energy (Fig. \ref{fig1}). Anisotropic resistivity, $AR$, measures this difference in response function.
\\
We have calculated the anisotropic resistivity by numerical solution of the semiclassical Boltzmann equation in the presence of the impurities in the graphene. Calculation within the Dirac point approximation shows that there is no anisotropic resistivity for this type of the effective Hamiltonian ($AR=0$).
\\
Regarding the value of the Fermi energy two distinct regimes could be recognized. As it was shown in Fig. \ref{fig2}(a) there are two different type of the Fermi curves corresponding to these regimes. At low energies when $\epsilon_F<t$ the Fermi curve of the system appears as separated islands in which by increasing the Fermi energy, $\epsilon_F$, the circular Fermi cross section changes into the trigonal one (Fig. \ref{fig2}(a)). This range of Fermi energies which includes the circular Dirac cones up to the trigonal Fermi curves has been called trigonal warping regime. Meanwhile at high Fermi energies, when $\epsilon_F>t$ Fermi curve is given by a closed hexagonal or closed smooth loop depending on the value of the Fermi energy (Fig. \ref{fig2}(b)). This case could be called connected Fermi curve regime.
\\
In the limit of the elastic scatterings in a single scattering process or even in a series of multiple sequential scatterings the initial and final states in k-space contributing in scattering should lie on the given Fermi curve. Meanwhile the form factor of the initial and final pseudospin states determines the amplitude of scattering between these states. The dependence of the scattering rates on the form factors introduces the contribution of the scattering matrix in the anisotropic resistivity of the system. As mentioned before there is also another contribution in the anisotropic resistivity of the system which characterizes by the anisotropy of the band energy. From the numerical point of view this contribution traces back to the Dirac delta function included in the expression: $f_\lambda(\phi ,\theta )={f}_{0}-e\varepsilon v_\lambda(-{{\partial }_{\epsilon }}{{f}_{0 \lambda}})(a_\lambda(\phi )\cos (\theta )+b_\lambda(\phi )\sin (\theta ))$ (where $-{{\partial }_{\epsilon }}{{f}_{0 \lambda}}=\delta(\epsilon_k-\epsilon_F)$) which selects the elastic scatterings among the other type of transitions. Meanwhile it should be noted that the Fermi velocity, which can be obtained directly from the band energy, ${{v}_{i}(\vec{k}_F)}=\frac{\partial E(k)}{\hbar \partial {{k}_{i}}}|_{k=k_F}$, contributes in the band-dependent anisotropy as well.
\\
As it was shown in Fig. \ref{fig3} when we restrict ourselves to the low Fermi energies of the trigonal wrapping regime i.e. when $\epsilon_F<<t$, known as Dirac approximation limit. In this limit Fermi curves appear as isotropic circles and anisotropic resistivity identically vanishes. At higher energies of this regime Fermi circles continuously come to change into the anisotropic trigonal curves. As mentioned before initial and final states in k-space which contributing in the elastic scattering process lie on a Fermi curve therefore in the trigonal warping regime in which the scattering cannot cover all of the possible orientations of the wave number, $k$, and anisotropy could have an accountable contribution in the anisotropic resistivity of the monolayer graphene. In this regime anisotropic resistivity characterizes by the sharp jumps as indicated in Fig. \ref{fig3}. Almost  all of these sharp peaks are positive in this range of Fermi energies. A sharp peak has been found at the transition point the trigonal warping regime.
\\
On the other hand at the limit of high Fermi energies i.e. in the connected Fermi curve regime Fermi curve appears as a continues hexagonal-like shape where increasing the Fermi energy deforms this shape into the circular or nearly circular Fermi curve. Therefore in this case all of the orientations in $k$-space are possible, however, except at low energies in which the Fermi curve is hexagonal-like the anisotropy of the band energy at high energies is really low for circular Fermi curves. Meanwhile as indicated in Fig. \ref{fig3} anisotropic resistivity oscillates by increasing the Fermi energy and the anisotropic resistivity of the system is absolutely negative in the range of Fermi energies corresponding to this regime. Since as mentioned before the contribution of band anisotropy is really low (except at limit of low energies) therefore it can be inferred that the anisotropy of their scattering has the main contribution in the AR. The dependence of the scattering rates on the form factors leads to these rapid oscillations in the AR. These oscillations could not be regarded as a contribution of the band anisotropy. This is due to the fact that the band anisotropy is less significant especially at low energies and more evidently the change of the Fermi curve by increasing the Fermi energy takes place very smoothly.
\\
Band anisotropy in the connected Fermi curve regime should be very low since as depicted in Fig. \ref{fig2} (b) the band energy is nearly circular at high energies. At first look it seems that trigonal warping could generate high anisotropic resistivity, however, results of the current study shows that the anisotropy introduced by the scattering matrix is of significant importance.
\\
\\
It is interesting to know that the different behavior of the anisotropic resistivity in these two regimes can be employed for the determination of the hopping amplitude, $t$. Since we have demonstrated that the transition point between these two regimes is $\epsilon_F=t$. Therefore full electric measurements of the conductivity in $x$ and $y$ directions could determine the nearest-neighbor hopping amplitude in monolayer graphene. This electric measurement of the hopping amplitude could be realized by determination of the transition point in the anisotropic resistivity curvature when depicted as a function of the Fermi energy (Fig. \ref{fig3}). At this point anisotropic resistivity undergoes a sign change after a sharp positive peak and starts to oscillate rapidly with Fermi energy (Fig. \ref{fig3}). It should be noted that the Fermi energy could also be controlled by an electric setup in which an external gate voltage controls the density of carriers and the amount of the Fermi energy. The above discussion would be fully fall down in the presence of the inelastic scatterings when the energy levels of the initial and final states of a single scattering is not the same and scattering rate gets more contribution in the anisotropic resistivity in both of these regimes.
\section{Conclusion}
In the current study we have shown that, the anisotropic resistivity in the graphene shows quite different behavior in two different regimes which were identified by ratio of the Fermi energy with respect to the hopping amplitude ($\epsilon_F/t$). At each of these regimes the functionality of anisotropic resistivity changes at the transition point of these two regimes. Results of the current study could be employed for determination of the hopping amplitude by full electric measurements of the conductivity in different directions.
\\

\bibliographystyle{elsarticle-num}
\bibliography{ref}

\begin{thebibliography}{10}
\expandafter\ifx\csname url\endcsname\relax
  \def\url#1{\texttt{#1}}\fi
\expandafter\ifx\csname urlprefix\endcsname\relax\def\urlprefix{URL }\fi
\expandafter\ifx\csname href\endcsname\relax
  \def\href#1#2{#2} \def\path#1{#1}\fi

\bibitem{1}
K.~S. Novoselov, D.~Jiang, F.Schedin, T.~J. Booth, V.~V. Khotkevich, S.~V.
  Morozov, A.~K. Geim, Natl. Acad.Sci. USA 102 (2005) 10451.

\bibitem{2}
K.~S. Novoselov, A.~K. Geim, S.~V. Morozov, D.~Jiang, Y.~Zhang, S.~V. Dubonos,
  I.~V. Grigorieva, , A.~A. Firsov, Science 306 (2004) 666.

\bibitem{14}
P.~R. Wallace, Phys. Rev 71 (1947) 622.

\bibitem{15}
J.~C. Slonczewski, P.~R. Weiss, Phys. Rev 109 (1958) 272.

\bibitem{16}
J.~W. McClure, Phys. Rev 108 (1957) 612.

\bibitem{17}
D.~DiVincenzo, E.~Mele, Phys. Rev. B 29 (1984) 1685.

\bibitem{18}
G.~W. Semeno, Phys. Rev. Lett 53 (1984) 2449.

\bibitem{33}
A.~H. Castro-Neto, F.~Guinea, N.~M.~R. Peres, K.~S. Novoselov, A.~K. Geim, Rev.
  Mod. Phys 81 (2009) 109.

\bibitem{35}
L.~Brey, H.~Fertig, Phys. Rev. B 73 (2006) 195408.

\bibitem{36}
L.~Chico, L.~X. Benedict, S.~G. Louie, M.~L. Cohen, Phys. Rev. B 54 (1996)
  2600.

\bibitem{38}
H.~Santos, L.~Chicod, L.~Brey, Phys. Rev. Lett 103 (2009) 086801.

\bibitem{39}
SonY-WCohen, M.~L. Loui, Nature (London) 444 (2006) 347.

\bibitem{40}
W.~L. Wang, S.~Meng, E.~Kaxiras, Nano Lett 8 (2008) 241.

\bibitem{41}
O.~V. Yazyev, M.~I. Katsnelson, Phys. Rev. Lett 100 (2008) 047209.

\bibitem{42}
L.~Yang, C.~H. Park, Y.~W. Son, S.~G. Loui, Phys. Rev.Lett. 99 (2007) 186801.

\bibitem{44}
O.~Hod, V.~Barone, J.~E. Perlata, G.~E. Scuseria, Nano Lett. 7 (2007) 2295.

\bibitem{45}
O.~Hod, V.~Barone, G.~E. Scuseria, Phys Rev B 77 (2008) 035411.

\bibitem{46}
O.~Hod, J.~E. Perlata, G.~E. Scuseria, Phys. Rev. B 76 (2007) 23340111.

\bibitem{47}
V.~Barone, O.~Hod, G.~E. Scuseria, Nano Lett 6 (2006) 2748.

\bibitem{48}
B.~Huang, F.~Liu, J.~Wu, G.~W. B-LDuan, Phys. Rev.B 77 (2008) 153411.

\bibitem{Chen}
Z.~Chen, Y.-M. Lin, M.~J. Rooks, P.~Avouris, Graphene nano-ribbon electronics,
  Physica E: Low-dimensional Systems and Nanostructures 40~(2) (2007) 228 --
  232.

\bibitem{23}
K.~Ziegler, Phys. Rev. Lett 97 (2006) 266802.

\bibitem{24}
V.~P. Gusynin, S.~G. Sharapov, Phys. Rev. Lett 95 (2005) 146801.

\bibitem{25}
N.~M.~R. Peres, F.~Guinea, A.~H.~C. Neto, Phys.Rev. B 73 (2006) 125411.

\bibitem{26}
M.~I. Katsnelson, Eur. Phys. J. B 51 (2006) 157.

\bibitem{27}
P.~M. Ostrovsky, I.~V. Gornyi, A.~D. Mirlin, Phys. Rev.B 74 (2006) 235443.

\bibitem{28}
J.~Tworzydlo, et~al, Phys. Rev. Lett 96 (2006) 246802.

\bibitem{10}
I.~L. Aleiner, K.~B. Efetov, Phys. Rev. Lett 97 (2006) 236801.

\bibitem{11}
H.~Ajiki, T.~Ando, J. Phys. Soc. Jpn 65 (1996) 505.

\bibitem{13}
T.~Ando, T.~Nakanishi, R.~Saito, J. Phys. Soc. Jpn 67 (1998) 2857.

\bibitem{29}
F.~Guinea, A.~H.~C. Neto, N.~M.~R. Peres, Phys.Rev. B 73 (2006) 245426.

\bibitem{30}
S.~Latil, L.~Henrard, Phys. Rev. Lett 97 (2006) 036803.

\bibitem{trig1}
K.~Kechedzhi, V.~I. Fal’ko, E.~McCann, , B.~L. Altshuler, Phys. Rev. Lett. 98
  (2007) 176806.

\bibitem{trig2}
M.~Koshino, E.~McCann, Phys. Rev. B 80 (2009) 165409.

\bibitem{trig3}
R.~Saito, G.~Dresselhaus, M.~S. Dresselhaus, Phys. Rev. B 61 (2000) 2981.

\bibitem{12}
K.~V\'{y}born\'{y}, A.~A. Kovalev, J.~Sinova, Jungwirth, Phys. Rev. B 26 (2008)
  0454271.

\end{thebibliography}







\end{document}